\documentclass[aps,floatfix,onecolumn,showpacs]{revtex4}
\usepackage{graphicx}
\usepackage{graphics,epsfig}
\usepackage{amssymb}

\def \H{{\cal H}}

\begin{document}

\title{Generalised perturbation equations in bouncing cosmologies}

\author{Antonio Cardoso}
\email{antonio.cardoso@port.ac.uk} \affiliation{Institute of
Cosmology \& Gravitation, University of Portsmouth,
Portsmouth~PO1~2EG, UK}

\author{David Wands}
\email{david.wands@port.ac.uk} \affiliation{Institute of Cosmology
\& Gravitation, University of Portsmouth, Portsmouth~PO1~2EG, UK}

\date{\today}

\begin{abstract}

We consider linear perturbation equations for long-wavelength scalar
metric perturbations in generalised gravity, applicable to
non-singular cosmological models including a bounce from collapse to
expansion in the very early universe. We present the general form
for the perturbation equations which follows from requiring that the
inhomogeneous universe on large scales obeys the same local
equations as the homogeneous Friedmann-Robertson-Walker background
cosmology (the separate universes approach). In a
pseudo-longitudinal gauge this becomes a homogeneous second-order
differential equation for adiabatic perturbations, which reduces to
the usual equation for the longitudinal gauge metric perturbation in
general relativity with vanishing anisotropic stress. As an
application we show that the scale-invariant spectrum of
perturbations in the longitudinal gauge generated during an
ekpyrotic collapse are not transferred to the growing mode of
adiabatic density perturbations in the expanding phase in a simple
bounce model.

\end{abstract}

\pacs{04.50.Kd, 98.80.Cq \hfill arXiv:0801.1667}

\maketitle

\section{Introduction}

An inflationary expansion in the very early universe can produce the
approximately scale-invariant spectrum of adiabatic density
perturbations observed in the cosmic microwave
background~\cite{Spergel:2006hy}, but recently there has been a
debate on whether this spectrum could also be generated in
cosmological models where the hot big bang phase is preceded by a
collapse phase. Among the proposed models are the pre-big
bang~\cite{Gasperini:1992em},
ekpyrotic~\cite{Khoury:2001wf,Lyth:2001pf}, and pressureless
collapse models~\cite{Wands:1998yp,Finelli:2001sr}.
%
%
Although many of the models seek to embed the four-dimensional
cosmology within a higher dimensional theory, most of the
quantitative calculations are done within an effective
four-dimensional theory which reduces to general relativity (GR) at
sufficiently low energies or late times. However, some modifications
to the dynamical equations are required at high energies, close to
the bounce, to avoid the singularities present in standard
cosmological solutions in GR. This has lead to confusion in the
literature, as different modifications appear to give different
results for the propagation of perturbations through the bounce. In
particular there are conflicting claims as to whether the dominant
mode in the longitudinal gauge metric potential $\Psi$, which has an
almost scale-invariant spectrum during an ekpyrotic collapse, could
in principle give rise to a scale-invariant spectrum in the growing
mode of adiabatic density perturbations after the
bounce~\cite{Lyth:2001pf,Durrer:2002jn,Cartier:2003jz,Bozza:2005qg,Copeland:2006tn,Alexander:2007zm}.


One approach which has been taken to analyze the transfer of metric
fluctuations through the bounce from the collapsing to the expanding
phase is to evolve the perturbations through an instantaneous
transition along a space-like
hypersurface~\cite{Hwang:1991an,Deruelle:1995kd}, using the analog
of the Israel junction conditions~\cite{Israel:1966rt}, which
describe the matching of two solutions of the Einstein equations
along a time-like hypersurface. This approach was discussed in
Ref.~\cite{Durrer:2002jn}, which suggested that generic matching
conditions would mix growing and decaying modes across the singular
hypersurface, but in Ref.~\cite{Copeland:2006tn} it was argued that
no such mixing occurs for adiabatic perturbations.
Another approach is to study the evolution of perturbations through
a non-singular bounce. This can be done by adding higher order terms
in the gravitational
action~\cite{Cartier:2001is,Tsujikawa:2002qc,Gasperini:2003pb}, but
the resulting field equations are complex and model-dependent.


It is tempting to seek a simpler prescription. In a recent work,
Alexander \textit{et al}~\cite{Alexander:2007zm} assumed that the
physics responsible for the bounce does not affect the form of the
evolution equation for the perturbations, and the only effect of
high-energy modifications at the bounce is to modify the background
Hubble rate. They choose a specific cosmological background which
leads to exactly soluble equations, and in which one finds that the
dominant mode of $\Psi$ during an ekpyrotic collapse couples to the
growing mode of density perturbations after the bounce. However, we
shall show that this approach implicitly introduces a non-adiabatic
component in the perturbations proportional to $\Psi$ on large
scales.

Instead we derive a generalised equation for $\Psi$ describing the
evolution of adiabatic perturbations in the long-wavelength limit
for a generalised Friedmann-Robertson-Walker (FRW) background, which
reduces to the standard equation when the background is general
relativity. This equation is derived by requiring that there exists
a long-wavelength limit in which the evolution of the perturbed
universe is the same as that of the FRW background. That is, we
assume that the same physics applies to long-wavelength
perturbations as applies to the homogeneous background
cosmology~\cite{Salopek:1990jq,Wands:2000dp,Bertschinger:2006aw}. In
order to obtain a closed second-order evolution equation for a
single metric perturbation we use the gauge freedom to work in a
pseudo-longitudinal gauge (defined in an Appendix) without using the
Einstein field equations. Adopting the same \textit{ansatz} for the
background Hubble rate as in~\cite{Alexander:2007zm} we find that
the dominant mode of $\Psi$ during an ekpyrotic collapse remains
decoupled from the growing mode of density perturbations after the
bounce.

Finally, we will comment on the relation of our work to previous
discussions of generalised perturbation equations on large
scales~\cite{Wands:2000dp,Bertschinger:2006aw,Starobinsky:2001xq}.
In particular we identify a generalization of the curvature
perturbation on uniform density hypersurfaces, $\zeta$, and show
that this quantity is constant in the long-wavelength limit for
adiabatic perturbations, even if the gravitational field equations
are modified and local conservation of energy is violated.

\section{Generalised equations for the perturbations}

We consider a gravitational theory where homogeneous and isotropic
spacetimes obey a Friedmann-type constraint equation, determining
the expansion rate of comoving worldlines, $\theta$, and an equation
for its evolution with respect to the proper time, $\tau$, along
these worldlines,
\begin{eqnarray}\label{perteq1}
\theta^2&=&3 f \,, \\
\frac{d}{d\tau}\theta&=&-\frac{3}{2} g \,. \label{perteq2}
\end{eqnarray}
For example, in loop quantum cosmology a modified effective
Friedmann equation (1) can be derived where $f=f(\rho)$ that leads
to a cosmological bounce (e.g.~\cite{Singh:2006im}), and in
Cardassian models~\cite{Freese:2002sq} $f(\rho) \propto
\rho+C\rho^n$ have been investigated. In both these examples local
energy conservation along comoving worldlines then fixes the form of
$g(\rho,p)$. For simplicity we neglect spatial curvature, which we
expect to be negligible in the early universe in any case.

In general relativity we then have $f=8\pi G\rho$ and $g=8\pi
G(\rho+p)$, where $G$ is Newton's constant, $\rho$ is the energy
density and $p$ the isotropic pressure of all types of
energy-momentum in the universe. More generally, one can always
define an effective energy-momentum tensor such that the Einstein
tensor $G_{\mu\nu}=8\pi GT_{\mu\nu}^{\rm{eff}}$. {}From
Eqs.~(\ref{perteq1}) and (\ref{perteq2}) we can identify an
effective density and pressure:
\begin{equation}
 \label{eff}
\rho^{\rm{eff}} \equiv \frac{f}{8\pi G} \,, \quad p^{\rm{eff}}
\equiv \frac{g-f}{8\pi G} \,.
\end{equation}
Conservation of the Einstein tensor, $\nabla^\mu G_{\mu\nu}=0$, then
requires conservation of the effective energy-momentum tensor, which
implies
\begin{equation}
\frac{d}{d\tau}\rho^{\rm{eff}} =
-\theta(\rho^{\rm{eff}}+p^{\rm{eff}}) \,,
\end{equation}
or equivalently, from Eqs.~(\ref{perteq1}) and (\ref{perteq2}),
\begin{equation}
\frac{d}{d\tau}f = -\theta g \,.
\end{equation}
However, in the following we will allow $f$ and $g$ to be arbitrary
functions of energy, pressure or other variables.
In particular we will not assume energy conservation.

We will assume that the spacetime that evolves into our observable
Universe can be described by small perturbations about a spatially
homogeneous and isotropic spacetime, as in the standard hot big bang
model. In particular we will consider scalar perturbations about a
spatially flat Friedmann-Robertson-Walker (FRW) metric. The linearly
perturbed line-element is given
by~\cite{Mukhanov:1990me,Malik:2001rm}
\begin{equation}\label{metric}
ds^2=a^2[-(1+2\phi)d\eta^2+2B,_idx^id\eta+\{(1-2\psi)\delta_{ij}+2E,_{ij}\}dx^idx^j]
\,,
\end{equation}
where $a(\eta)$ is the background scale factor and a comma denotes a
partial derivative with respect to the 3D spatial coordinates. In
general one should also consider vector and tensor metric
perturbations, but at linear order these decouple from the scalar
modes which describe the primordial density
perturbations~\cite{Bardeen:1980kt}.

There is a unit time-like vector field orthogonal to constant-$\eta$
spatial hypersurfaces~\cite{Kodama:1985bj,Malik:2001rm},
\begin{equation}\label{orthovector}
N^{\mu}=\frac{1}{a}(1-\phi,-B,^i) \,,
\end{equation}
whose expansion rate is given by
\begin{equation}\label{theta}
\theta=N^{\mu};_{\mu} \,,
\end{equation}
where a semicolon denotes a covariant 4D derivative. In terms of the
scalar metric perturbations we have
\begin{equation}\label{thetaeq}
\theta=3\frac{a'}{a^2}(1-\phi)-\frac{3}{a}\psi'+\frac{1}{a}\nabla^2\sigma
\,,
\end{equation}
where a prime denotes a derivative with respect to the conformal
time $\eta$, $\nabla^2$ denotes the 3D spatial Laplacian and
$\nabla^2\sigma$ is the shear, with
\begin{equation}\label{sigma}
\sigma=E'-B \,.
\end{equation}
At zeroth-order the shear vanishes and the background expansion rate
is $\theta_0=3\H/a$, where $\H\equiv a'/a$ is the conformal Hubble
parameter.

For the zeroth-order homogeneous (FRW) background the
equations~(\ref{perteq1}) and~(\ref{perteq2}) can be written as
\begin{eqnarray}\label{backeq1}
\H^2&=&\frac{a^2}{3} f_0 \,, \\
\H'-\H^2&=&-\frac{a^2}{2} g_0 \,. \label{backeq2}
\end{eqnarray}
%

Adopting the separate universes viewpoint~\cite{Wands:2000dp}, the
same equations~(\ref{perteq1}) and~(\ref{perteq2}) can be used to
describe the local evolution in an inhomogeneous universe if we work
in the long-wavelength limit where physical length scales are much
larger than the time scale set by the expansion rate, $\lambda\gg
\theta^{-1}$.
This is also sometimes called the ultra-local
approximation~\cite{Erickson:2006wc}.
%
Of course the expansion time $\theta^{-1}$ diverges at any
cosmological bounce ($\theta=0$) in the very early universe, but the
bounce itself must have a characteristic time scale associated with
very high energies and thus very short time scales.

We then can apply Eqs.~(\ref{perteq1}) and~(\ref{perteq2}) where we
take $f=f_0(\eta)+\delta f(\eta,{\bf x})$ and $g=g_0(\eta)+\delta
g(\eta,{\bf x})$ and the local expansion rate is given, to
first-order, by Eq.~(\ref{thetaeq}).
%
%
Neglecting all spatial gradients, we can then write the first-order
equations in terms of the lapse function $\phi$, its derivative, the
curvature perturbation $\psi$ and its first and second derivatives,
\begin{eqnarray}\label{perteqmetric1}
3\H(\psi'+\H\phi)&=&-\frac{a^2}{2}\delta f \,, \\
\psi''-\H\psi'+\H\phi'+2(\H'-\H^2)\phi&=&\frac{a^2}{2}\delta g \,.
\label{perteqmetric2}
\end{eqnarray}
Note that these equations are independent of two of the scalar
metric perturbations, $B$ and $E$ in Eq.~(\ref{metric}), which
determine the anisotropic shear~(\ref{sigma}), which vanishes in
this long-wavelength limit.

For adiabatic perturbations on large scales different patches of the
inhomogeneous universe follow the same trajectory in phase space,
and the adiabatic perturbations correspond to a perturbation back or
forward with respect to the homogeneous background along this
trajectory~\cite{Wands:2000dp}. In this case the hypersurfaces of
uniform-$\theta$ and uniform-$(d\theta/d\tau)$ coincide. To
first-order this requires $\delta g/g_0'=\delta f/f_0'$.

More generally, we can write any perturbation $\delta g$ as a sum of
its adiabatic and non-adiabatic parts,
\begin{equation}\label{deltag}
\delta g=\frac{g_0'}{f_0'}\delta f+\delta g_{\text{nad}} \,,
\end{equation}
where $\delta g_{\text{nad}}$ is automatically gauge-invariant.
Indeed, if we identify $f$ with an effective density and $g-f$ with
an effective pressure, then $\delta g_{\rm nad}=8\pi G[\delta p^{\rm
eff}-({p_0^{\rm eff}}'/{\rho_0^{\rm eff}}')\delta \rho^{\rm
eff}]=8\pi G\delta p_{\rm nad}^{\rm eff}$. If we assume $f=f(\rho)$
in Eq.~(\ref{perteq1}) and impose energy conservation, so that
$d\rho/d\tau+\theta(\rho+p)=0$ along comoving worldlines, then we
would require from Eq.~(\ref{perteq2}) that $g=(df/d\rho)(\rho+p)$
and then $\delta g_{\rm nad}=(df/d\rho)\delta p_{\rm nad}$.

Using Eqs.~(\ref{deltag}),~(\ref{backeq1}) and~(\ref{backeq2}), we
have from Eqs.~(\ref{perteqmetric1}) and~(\ref{perteqmetric2}) that
\begin{equation}\label{psiphieq}
\psi''+\frac{3\H\H'-\H''-\H^3}{\H'-\H^2}\psi'+\frac{\H\H'-\H^3}{\H'-\H^2}\phi'+\frac{2\H'^2-\H\H''}{\H'-\H^2}\phi=\frac{a^2}{2}\delta
g_{\text{nad}} \,.
\end{equation}

Equation~(\ref{psiphieq}) includes the two gauge-dependent metric
perturbations $\psi$ and $\phi$. However, we still have a choice of
temporal gauge. If we choose a pseudo-longitudinal gauge in which
$\psi=\phi\equiv\Psi$ (see Appendix~A) we find the following
evolution equation:
\begin{equation}\label{Psieq}
\Psi''+\frac{4\H\H'-\H''-2\H^3}{\H'-\H^2}\Psi'+\frac{2\H'^2-\H\H''}{\H'-\H^2}\Psi=\frac{a^2}{2}\delta
g_{\text{nad}} \,.
\end{equation}
For adiabatic perturbations the right-hand-side vanishes and we have
a homogeneous second-order evolution equation for $\Psi$.
We can solve this equation by quadratures to find the general
solution~\cite{Bertschinger:2006aw}
\begin{equation}
 \label{psisolution}
 \Psi = C \frac{\H}{a^2} + K \frac{\H}{a^2} \int
 \frac{a^2(\H'-\H^2)}{\H^2} \,
 d\eta \,,
\end{equation}
where $C$ and $K$ are constants of integration. This also has a
simpler form~\cite{Polarski:1992dq}:
\begin{equation}
 \label{psisolutionsimp}
 \Psi = C \frac{\H}{a^2} + K \left[-1+\frac{\H}{a^2} \int
 a^2 \,
 d\eta\right] \,,
\end{equation}
Although the differential equation~(\ref{Psieq}) has a singular
point when $\H'-\H^2=0$, the solution~(\ref{psisolutionsimp}) is
clearly regular through a bounce. We show in Appendix~A that $\Psi =
C \H/a^2$ is a solution on all scales in the pseudo-longitudinal
gauge, but the second term is only a solution in the long-wavelength
limit.


\subsection{General relativity limit}

For general relativity we know the full equations for linear
perturbations about a FRW metric. The background expansion rate
obeys the Friedmann constraint and evolution
equations,~(\ref{backeq1}) and~(\ref{backeq2}), with $f_0=8\pi
G\rho_0$ and $g_0=8\pi G(\rho_0+p_0)$, while we have the following
perturbation equations, including the spatial
gradients,~\cite{Mukhanov:1990me,Malik:2001rm}:
\begin{eqnarray}\label{perteqgr1}
3\H(\psi'+\H\phi)-\nabla^2\psi-\H\nabla^2\sigma&=&-4\pi Ga^2 \delta\rho \,, \\
\psi''+2\H\psi'+\H\phi'+(2\H'+\H^2)\phi&=&4\pi G a^2 (\delta
p+\frac{2}{3}\nabla^2\Pi) \,, \label{perteqgr2}
\end{eqnarray}
where $\Pi_{,ij}-(1/3)\delta_{ij}\nabla^2\Pi$ is the anisotropic
stress.
It is straightforward to see that Eqs.~(\ref{perteqmetric1})
and~(\ref{perteqmetric2}) reduce to Eqs.~(\ref{perteqgr1})
and~(\ref{perteqgr2}), if we neglect the spatial gradient terms, in
general relativity, where $\delta f = 8\pi G \delta\rho$ and $\delta
g=8\pi G(\delta\rho+\delta p)$.

General relativity also supplies a momentum constraint equation, and
an evolution equation for the anisotropic shear potential, $\sigma$.
In the longitudinal gauge, where $\sigma=0$, this becomes a
constraint equation which requires $\psi=\phi$ in the absence of
anisotropic stress, $\Pi=0$,~\cite{Mukhanov:1990me}.
From Eqs.~(\ref{perteqgr1}),~(\ref{perteqgr2}) and~(\ref{deltap}) we
obtain
\begin{equation}\label{Psieqgr}
\Psi''+3(1+c_{s}^2)\H\Psi'+[2\H'+(1+3c_s^2)\H^2-c_s^2\nabla^2]\Psi=4\pi
Ga^2\delta p_{\text{nad}} \,,
\end{equation}
where here $\Psi$ is the curvature perturbation in the longitudinal
gauge, and the non-adiabatic pressure perturbation is
\begin{equation}\label{deltap}
\delta p=c_s^2\delta\rho+\delta p_{\text{nad}} \,,
\end{equation}
where $c_{s}^2\equiv p_0'/\rho_0'$ is the adiabatic sound speed of
the matter.

Using Eqs.~(\ref{backeq1}) and~(\ref{backeq2}) it is easy to show
that in general relativity Eq.~(\ref{Psieq}) reduces to
Eq.~(\ref{Psieqgr}), neglecting the spatial gradients, with $\delta
g_{\text{nad}}=8\pi G\delta p_{\text{nad}}$.

More generally, if we use equations (\ref{eff}) to identify an
effective density and pressure on large scales, then one can show
that our generalised perturbation equation (\ref{Psieq}) can be
written in a ``general relativistic'' form
\begin{equation}
\Psi''+3(1+c_{s}^{2\rm{eff}})\H\Psi'+[2\H'+(1+3c_s^{2\rm{eff}})\H^2]\Psi=4\pi
Ga^2\delta p_{\text{nad}}^{\rm{eff}} \,,
\end{equation}
where the effective adiabatic sound speed is
\begin{equation}
c_s^{2\rm{eff}} =
\frac{p_0^{\rm{eff}\prime}}{\rho_0^{\rm{eff}\prime}} =
\frac{\H\H'+\H^3-\H''}{3\H(\H'-\H^2)} \,.
\end{equation}

\section{Application to a simple bouncing cosmology}

We will now consider the evolution of perturbations in the simple
bounce cosmology proposed by Alexander \textit{et al} in
Ref.~\cite{Alexander:2007zm}. They assumed a barotropic fluid in the
universe, with pressure $p=w\rho$, and adiabatic density
perturbations.
To model the bounce they assumed a specific \textit{ansatz} for the
background evolution, with the conformal Hubble rate given by
\begin{equation}\label{Hmodified}
\H=\frac{q\eta}{\eta^2+\eta_0^2} \,,
\end{equation}
where $q\equiv2/(1+3w)$, for which the scale factor has the solution
\begin{equation}\label{amodified}
a=a_i\left(\frac{\eta^2+\eta_0^2}{\eta_i^2+\eta_0^2}\right)^{\frac{q}{2}}
\,,
\end{equation}
where $\eta_i$ and $a_i$ denote the initial values of the conformal
time and scale factor, respectively. At early or late times,
$\vert\eta\vert\gg\eta_0$, we recover the general relativistic
background evolution
\begin{equation}\label{Hgr}
\H=\frac{q}{\eta} \,,
\end{equation}
but for $|\eta|\sim\eta_0$ we require some unspecified modification
of the gravitational field equations to produce the modified
expansion given in Eq.~(\ref{Hmodified}).

In the general relativistic limit (at early and late times,
$|\eta|\gg\eta_0$) we can use either the GR perturbation
equation~(\ref{Psieqgr}) or its generalisation,~(\ref{Psieq}), to
obtain
\begin{equation}\label{PsieqgrHgr}
\Psi^{\prime\prime}+\frac{6(1+w)}{\eta(1+3w)}\Psi^{\prime}=0 \,
\end{equation}
for adiabatic perturbations in the long-wavelength limit.
This has the analytic solution
\begin{equation}\label{solPsieqgrHgr}
\Psi=D_{\pm}+S_{\pm}(\pm\eta)^{-2\nu} \,,
\end{equation}
where $\nu\equiv(5+3w)/[2(1+3w)]$, $D_{\pm}$ and $S_{\pm}$ are
constants of integration and the $+$ and $-$ signs correspond to the
expanding and contracting phases, respectively.

In the ekpyrotic limit, $w\rightarrow\infty$, we obtain
\begin{equation}\label{solPsieqgrHgrwinfty}
\Psi=D_{\pm}+S_{\pm}(\pm\eta)^{-1} \,.
\end{equation}
Normalising to initial vacuum fluctuations in the ekpyrotic model
gives a scale-invariant spectrum of fluctuations in the growing mode
approaching the bounce, $\Psi\sim S_-(-\eta)^{-1}$, while the
constant mode, $D_-$, has a steep blue spectrum~\cite{Lyth:2001pf}.
However, observations constrain the spectrum of perturbations in the
adiabatic growing mode in the expanding phase, $D_+$.


If we assume that the perturbations are adiabatic in the
long-wavelength limit, that is, $\delta g_{\text{nad}}=0$ in
Eq.~(\ref{Psieq}), we have
\begin{equation}\label{Psieqadiabatic}
\Psi''+\frac{4\H\H'-\H''-2\H^3}{\H'-\H^2}\Psi'+\frac{2\H'^2-\H\H''}{\H'-\H^2}\Psi=0
\,.
\end{equation}

In the $w\rightarrow\infty$ limit, substituting in
Eq.~(\ref{Hmodified}) for $\H(\eta)$, Eq.~(\ref{Psieqadiabatic})
reduces to
\begin{equation}\label{PsieqHmodified}
\Psi^{\prime\prime}+\frac{2\eta^3-6\eta_0^2\eta}{\eta^4-\eta_0^4}\Psi^{\prime}=0
\,.
\end{equation}
This has the solution
\begin{equation}\label{solPsieqHmodified}
\Psi=C_1+C_2\frac{\eta}{\eta^2+\eta_0^2} \,,
\end{equation}
where $C_1$ and $C_2$ are constants. Matching the solution in
Eq.~(\ref{solPsieqHmodified}) to the solution in
Eq.~(\ref{solPsieqgrHgrwinfty}) at both early and late times,
$\eta\rightarrow\pm\infty$, we get
\begin{equation}\label{matching}
D_{+}=D_{-} \quad \text{and} \quad
S_{+}= - S_{-} \,,
\end{equation}
therefore we see that there is no mixing between the different modes
in the early and late time limits. In particular the scale-invariant
spectrum during an ekpyrotic collapse, $S_-$, couples only to the
decaying mode $S_+$ in the expanding phase, and the adiabatic
growing mode in the expanding phase, $D_+$, has a steep blue
spectrum.


This contrasts with the results presented
in~\cite{Alexander:2007zm}, where it was assumed that, although the
background evolution is modified, Eq.~(\ref{Hmodified}), the
perturbations are governed by the general relativistic
equation~(\ref{Psieqgr}). Neglecting spatial gradients and setting
$\delta p_{\text{nad}}=0$ they obtain the following solution when
$w\rightarrow\infty$
\begin{equation}\label{solPsieqgrHmodified}
\Psi=B_1+B_2\arctan\left(\frac{\eta}{\eta_0}\right) \,,
\end{equation}
where $B_1$ and $B_2$ are constants. Matching this solution to the
solution in Eq.~(\ref{solPsieqgrHgrwinfty}), at both early and late
times, $\eta\rightarrow\pm\infty$, they obtain
\begin{equation}\label{matchinggrHmodified}
D_{+}=D_{-}+\frac{\pi}{\eta_0}S_{-} \quad \text{and} \quad
S_{+}=-S_{-} \,.
\end{equation}
In this case the mode $D_{+}$ receives a contribution from both
modes in the collapsing phase.

Comparing our Eq.~(\ref{Psieq}) with Eq.~(\ref{Psieqgr}) we see
that, by modifying the background evolution but using the unmodified
GR perturbation equation, the authors of
Ref.~\cite{Alexander:2007zm} implicitly introduced a non-adiabatic
perturbation in the evolution Eq.~(\ref{Psieq}),
\begin{equation}\label{deltaggrHmodified}
\delta
g_{\text{nad}}=-\frac{2}{a^2}\left[3\H(1+c_s^2)+\frac{\H''+2\H^3-4\H\H'}{\H'-\H^2}\right](\Psi'+\H\Psi)
\,.
\end{equation}
Note that Alexander {\em et al}~\cite{Alexander:2007zm} show that
gradient terms are not important on large scales close to the bounce
($k\eta_0\ll1$). However, by using the general relativistic equation
they have implicitly introduced a non-adiabatic perturbation,
$\delta g_{\text{nad}}$, proportional to the longitudinal gauge
metric potential $\Psi$, and this is why the $D_+$ mode in the
expanding phase receives a contribution from $S_-$ which would not
otherwise contribute to the growing mode in the expanding phase.

Mode mixing is certainly possible, indeed it happens in general
relativity, but only on finite scales due to $\nabla^2\Psi\neq0$
(see e.g.,~\cite{Gordon:2000hv}). But mixing occurs due to spatial
gradients and hence is suppressed on large scales.
Of course we cannot say that there is no theory in which a
non-adiabatic perturbation of the form given in
Eq.~(\ref{deltaggrHmodified}) could occur, but one would need a
physical model for the appearance of non-adiabatic source terms
unsuppressed on large scales (e.g., a non-adiabatic perturbation of
a multi-component system).
We have shown that by assuming only adiabatic perturbations on large
scales, derived from linear perturbations about the background
solutions to Eqs.~(\ref{perteq1}) and (\ref{perteq2}), no such
mixing occurs.

\section{Conclusions}

In this paper we have derived a generalised equation for the
evolution of scalar metric perturbations in the long-wavelength
limit, for a generalised Friedmann-Robertson-Walker (FRW)
background, which reduces to the standard equation for the
longitudinal metric perturbation in general relativity.

Our results are consistent with previous
work~\cite{Wands:2000dp,Starobinsky:2001xq} which pointed out that
the curvature perturbation on uniform-density
hypersurfaces~\cite{Bardeen:1983qw,Wands:2000dp},
\begin{equation}
\zeta \equiv - \psi - \frac{\H}{\rho_0'}\delta\rho \,,
\end{equation}
is conserved for adiabatic perturbations on large scales assuming
only local conservation of energy (see
also~\cite{Lyth:2003im,Lyth:2004gb,Creminelli:2004jg}).
We can define a generalization
\begin{equation}\label{zeta}
\zeta_f\equiv-\psi-\frac{\H}{f_0'}\delta f \,
\end{equation}
which is the gauge-invariant definition of the curvature
perturbation, $-\psi$, on uniform-expansion hypersurfaces, where
$\delta f=0$.
Using Eqs.~(\ref{backeq1}) and~(\ref{perteqmetric1}) we can write
\begin{equation}\label{zetapsi}
\zeta_f =
\frac{1}{\H'-\H^2}\left[\H\psi'+(\H^2-\H')\psi+\H^2\phi\right] \,.
\end{equation}
%
%
%
In general relativity the uniform-density and uniform-expansion
hypersurfaces coincide in the long-wavelength limit and hence
$\zeta_f=\zeta$.
Using Eqs.~(\ref{perteqmetric1}) and~(\ref{perteqmetric2}) for the
evolution of perturbations on large scales we obtain
\begin{equation}\label{zetaprime}
\zeta_f' = \frac{\H}{\H'-\H^2}\frac{a^2}{2}\delta g_{\text{nad}} \,.
\end{equation}
We see that $\zeta_f$ is constant in the large-scale limit for
modified gravitational field equations, even allowing for
non-conservation of energy, if the perturbations are adiabatic,
i.e., $\delta g_{\text{nad}}=0$ in Eq.~(\ref{deltag}).

We see that the growing mode solution (in an expanding universe) for
the pseudo-longitudinal gauge perturbation, $\Psi_+\propto K$ in
Eq.~(\ref{psisolution}), corresponds to $\zeta_f=K$, where $K$ is a
constant of integration. The decaying mode $\Psi_-=C\H/a^2$ does not
contribute to the curvature perturbation $\zeta_f$ and is the
residual gauge mode in the definition of the pseudo-longitudinal
gauge (see Appendix~A). The other physical degree of freedom in the
pseudo-longitudinal gauge resides in the anisotropic shear
potential, $\tilde\sigma$, which does not affect $\zeta_f$ or the
large-scale evolution equation~(\ref{Psieq}) in the long-wavelength
limit.

The same solution~(\ref{psisolution}) was given by
Bertschinger~\cite{Bertschinger:2006aw} in modified gravity for the
longitudinal gauge metric perturbation assuming that the trace-free
spatial part of the Einstein tensor vanishes (equivalent to assuming
vanishing anisotropic stress in GR). We have shown that it is
possible to use the solution~(\ref{psisolution}) for adiabatic
cosmological perturbations on large scales in a pseudo-longitudinal
gauge without making any assumption about the gravity theory, energy
conservation or anisotropic stress. The GR solution for the shear in
the pseudo-longitudinal gauge in the absence of anisotropic stress
(e.g., in a conventional hot big bang phase) is given by
$\tilde\sigma=\tilde{C}/a^2$, where $\tilde{C}$ is a constant of
integration. In this case we can perform a further gauge
transformation $\delta\eta_\ell=\tilde\sigma$ [see
Eq.~(\ref{transsigma})] to obtain the usual longitudinal gauge
curvature perturbation, $\Psi_\ell=\Psi+\tilde{C}\H/a^2$, which
shows that in the GR limit the longitudinal and pseudo-longitudinal
solutions differ only in the decaying mode, $\propto \H/a^2$.

We applied our formalism to adiabatic perturbations in a simple
cosmological bounce model, assuming a specific \textit{ansatz} for
the background evolution introduced in Ref.~\cite{Alexander:2007zm}.
We took the limit where the equation of state of the matter present
in the universe $w=p/\rho$ goes to infinity, corresponding to an
ekyprotic model. We conclude that the dominant mode of the curvature
perturbation after the bounce does not receive a contribution from
the growing mode in the collapse phase. This result is in contrast
with the result presented in Ref.~\cite{Alexander:2007zm}, where
they have assumed that, although the background evolution is
modified, the perturbations are governed by the same equation as in
general relativity. We have shown that such an approach is
equivalent to introducing a non-adiabatic component proportional to
$\Psi$ on large scales.

Since it is not possible to unambiguously separate perturbations
from the background in the large scale limit where spatial gradients
are negligible, we argue that if the background equations are
modified then the perturbation equations must also be modified in
the large scale limit. And Eq.~(\ref{Psieq}), not
Eq.~(\ref{Psieqgr}), is the consistent generalisation of the
evolution equation for long-wavelength linear perturbations given a
modified background solution for $\H(\eta)$.

The general nature of the perturbation equation~(\ref{Psieq}) and
solution~(\ref{psisolutionsimp}), derived only assuming generalised
gravity field equations of the form given in~(\ref{perteq1})
and~(\ref{perteq2}), implies that the growing mode of metric
perturbations~(\ref{solPsieqgrHgrwinfty}) in an ekpyrotic phase will
only contribute to the decaying mode in an expanding phase,
independently of the details of the bounce or the expanding phase.

\section*{Acknowledgements}
The authors are grateful to Alexei A. Starobinsky for drawing their
attention to Ref.~\cite{Polarski:1992dq}. AC is supported by FCT
(Portugal) PhD fellowship SFRH/BD/19853/2004. DW is supported by
STFC.

\appendix
\section{Pseudo-longitudinal gauge}

The metric perturbations $\phi$ and $\psi$ given in
Eq.~(\ref{metric}) and $\sigma$ defined in Eq.~(\ref{sigma}) are
gauge-dependent. Under a change of time coordinate
$\eta\to\eta+\delta\eta$ we have~\cite{Mukhanov:1990me,Malik:2001rm}
\begin{eqnarray}
\label{transphi}
 \phi &\to& \tilde\phi = \phi - \H\delta\eta - \delta\eta' \,,\\
\label{transpsi}
 \psi &\to& \tilde\psi = \psi + \H\delta\eta \,,\\
\label{transsigma}
 \sigma &\to& \tilde\sigma = \sigma - \delta\eta
\,.
\end{eqnarray}
Therefore we can always choose
\begin{equation}
 \label{pleta}
\delta\eta' + 2\H\delta\eta= \phi - \psi \,
\end{equation}
such that from Eqs.~(\ref{transphi}) and~(\ref{transpsi}) we have
$\tilde\psi=\tilde\phi\equiv\Psi$. This fixes the gauge up to an
arbitrary constant of integration $C({\bf x})$, which is similar to
what occurs in the synchronous gauge. Perturbations in this {\em
pseudo-longitudinal gauge} are given in terms of metric
perturbations in an arbitrary gauge by substituting
Eq.~(\ref{pleta}) in either Eq.~(\ref{transphi})
or~(\ref{transpsi}), to obtain
\begin{equation}
\Psi = \psi + \frac{\H}{a^2} \int a^2 (\phi-\psi) \, d\eta \,.
\end{equation}
We see that we can write the general solution for $\Psi$ as
$\Psi_+(\eta)+\Psi_-(\eta)$, where $\Psi_-(\eta)=C\H/a^2$ decays in
an expanding universe for $w>-1$.

In general relativity we have a field equation coming from the
trace-free part of the Einstein tensor which requires
\begin{equation}\label{Pi}
\psi-\phi = 8\pi G a^2 \Pi -\sigma' - 2h\sigma \,,
\end{equation}
where the scalar part of the anisotropic stress is given by
$\nabla_i\nabla_j\Pi-(1/3)\delta_{ij}\nabla^2\Pi$. In the
longitudinal or Newtonian gauge~\cite{Mukhanov:1990me} we have
$\sigma_\ell=0$ by definition, and hence $\psi_\ell=\phi_\ell$ in
the absence of anisotropic stress. More generally, we can define a
pseudo-longitudinal gauge in which $\tilde\psi=\tilde\phi$ and the
shear potential is given by
\begin{equation}
\tilde\sigma' + 2h\tilde\sigma = 8\pi G a^2 \Pi \,.
\end{equation}
Therefore, in the absence of anisotropic stress, $\Pi=0$, we have
$\tilde\sigma\propto a^{-2}$.

In generalised gravity we may not have a field equation of the same
form as Eq.~(\ref{Pi}), so we cannot in general identify the
pseudo-longitudinal gauge $\tilde\psi=\tilde\phi$ with a zero-shear
gauge. If we assume that there exists a general relativistic limit
in which the pseudo-longitudinal gauge reduces to the usual
longitudinal gauge (in the absence of anisotropic stress) then this
fixes the otherwise arbitrary constant of integration, $C({\bf x})$,
and fixes the gauge throughout.

\end{document}